\documentstyle[prl,aps,twocolumn]{revtex}

\newcommand{\R}{{\bf R}}
\newcommand{\eps}{\varepsilon}
\newcommand{\K}{{\cal K}}
\newcommand{\x}{{\bf r}}
\newcommand{\X}{{\bf X}}
\newcommand{\Tr}{{\rm Trace}\, }
\newcommand{\half}{\mbox{$\frac{1}{2}$}}
\newcommand{\E}{{\mathcal E}^{\rm GP}}
\newcommand{\Engp}{E^{\rm GP}}
\newcommand{\Enqm}{E^{\rm QM}}
\newcommand{\pgp}{\phi^{\rm GP}}
\newcommand{\k}{{\bf k}}

\draft
\begin{document}

\twocolumn[
\hsize\textwidth\columnwidth\hsize\csname@twocolumnfalse\endcsname

\title{Proof of Bose-Einstein Condensation for Dilute Trapped Gases}
\author{Elliott H. Lieb and Robert Seiringer\cite{leave}}
\address{Department of Physics, Jadwin Hall, Princeton University,
P.O. Box 708, Princeton, New Jersey 08544, USA}
\date{December 16, 2001}
\maketitle

\begin{abstract}
The ground state of bosonic atoms in a trap has been shown
experimentally to display Bose-Einstein condensation (BEC).  We
prove  this fact theoretically for bosons with two-body repulsive
interaction potentials in the dilute limit, starting from the basic
Schr\"odinger equation; the condensation is 100\% into the state
that minimizes the Gross-Pitaevskii energy functional.  This is
the first rigorous proof of BEC in a physically realistic,
continuum model.
\end{abstract}

\vfill  \twocolumn
\vskip.5pc ] \narrowtext

It is gratifying to see the experimental realization, in traps, of
the long-predicted Bose-Einstein condensation (BEC) of gases. From
the theoretical point of view, however, a rigorous demonstration
of this phenomenon -- starting from the many-body Hamiltonian of
interacting particles -- has not yet been achieved. In this letter
we provide such a rigorous justification  for the ground state of
2D or 3D bosons in a trap with repulsive pair potentials, and in
the well-defined limit (described below) in which the Gross-Pitaevskii
(GP) formula is applicable. It is the first proof of BEC for interacting
particles in a continuum (as distinct from lattice) model and in a
physically realistic situation.

The difficulty of the problem comes from the fact that BEC is not
a consequence of energy considerations alone. The correctness
\cite{LY98}  of Bogolubov's formula for the ground state energy
per particle, $e_0(\rho)$, of bosons at low density $\rho$, namely
$e_0(\rho) = 2\pi \hbar^2 \rho a /m$ (with $m=$ particle mass and  $a=$
scattering length of the pair potential) shows only that
`condensation' exists on local length scales. The same is true
\cite{LY01} in 2D, with Schick's formula \cite{Schick} $e_0(\rho)
= 2\pi \hbar^2 \rho  / (m |\ln (\rho a^2)|)$. Although it is convenient to
{\it assume} BEC in the derivation of $e_0(\rho)$,  these formulas for
$e_0(\rho)$ do not {\it prove} BEC. Indeed, in 1D the assumption
of BEC leads to a correct formula \cite{LL63} for $e_0(\rho)$, but
there is, presumably, no BEC in 1D ground states \cite{PS91}.

The results just mentioned are for homogeneous gases in the thermodynamic
limit. For traps, the GP formula is exact \cite{LSY00,LSY01} in the
limit, and one expects BEC into the GP 
function (instead of into the constant, or zero
momentum, function appropriate for the homogeneous gas). This is 
proved  in Theorem 1. In  the homogeneous case the BEC is not
100\%, even in the ground state. There is always some `depletion'. In
contrast, BEC in the GP limit  is 100\% because the $N\to \infty$
limit is different.

In the homogeneous case one fixes $a>0$ and takes $N\to \infty$
with $\rho = N/{\mathrm{volume}}$ fixed. For the GP limit one fixes
the external trap potential $V(\x)$ and fixes $Na$, the `effective
coupling constant', as $N\to \infty$.  A particular, academic
example of the trap is $V(\x)=0$ for $\x$ inside a unit cube and
$V(\x) =\infty $ otherwise. By scaling, one can relate this
special case to the homogeneous case and thereby compare the two
limits; one sees that the homogeneous case corresponds,
mathematically, to the trap case with this special $V$, but with
$N a ^3 =\rho a^3$ fixed as $N\to \infty$. Thus, BEC in the trap case 
is the easier of the two, 
reflecting the incompleteness of  BEC in the
homogeneous case. 
The lack of depletion in the GP limit is
consistent with $\rho a^3 \to 0$ and with  
Bogolubov theory. 

We now describe the setting more precisely. We concentrate on the
3D case, and comment on the generalization to 2D at the end of
this letter. The Hamiltonian for $N$ identical bosons in a trap
potential $V$, interacting via a pair potential $v$, is
\begin{equation}\label{ham}
H=\sum_{i=1}^N \left(-\Delta_i+V(\x_i)\right)+\sum_{1\leq i<j\leq
N} v(\x_i-\x_j).
\end{equation}
It acts on  symmetric functions of $N$ variables $\x_i \in \R^3$.
Units in which $\hbar^2/2m=1$ are used. We assume the trap
potential $V$ to be a locally bounded function, that tends to
infinity as $|\x|\to\infty$. The interaction potential $v$ is
assumed to be nonnegative, spherically symmetric, and have a
finite scattering length $a$. (For the definition of scattering
length, see \cite{LSY00}, \cite{LY01} or \cite{LY98}.) Note that we do not
demand $v$ to be locally integrable; it is allowed to  have a hard
core, which forces the wave functions to vanish whenever two
particles are close together. In the following, we want to let
$a$ vary with $N$, and we do this by scaling, i.e., we write
$v(\x)=v_1(\x/a)/a^2$, where $v_1$ has scattering length $1$, and
keep $v_1$ fixed when varying $a$.

The Gross-Pitaevskii functional is given by
$$
\E[\phi]=\int\left(|\nabla\phi(\x)|^2+V(\x) |\phi(\x)|^2 + g
|\phi(\x)|^4\right)d^3\x.
$$
The parameter $g$ is related to the scattering length of the interaction
potential appearing in (\ref{ham}) via
\begin{equation}
 g=4\pi N a.
\end{equation}
We denote
by $\pgp$  the minimizer of $\E$ under the normalization condition
$\int|\phi|^2=1$. Existence, uniqueness, and  some regularity
properties of $\pgp$ were proved in the appendix of \cite{LSY00}. In
particular, $\pgp$ is continuously differentiable and strictly
positive. Of course $\pgp$ depends on $g$, but we omit this dependence
for  simplicity of  notation.  For use later, we define the projector
\begin{equation}
P^{\rm GP}= |\pgp\rangle\langle \pgp|\ .
\end{equation}

It was shown in \cite{LSY00} (see also Theorem 2 below)  that, for
each  fixed $g$, the minimization of the GP functional correctly reproduces
the large $N$ asymptotics of the ground state energy and density of $H$ --
but no assertion about BEC in this limit was made in \cite{LSY00}.

BEC in $\Psi$, the (nonnegative and normalized) ground state of
$H$, refers to  the reduced one-particle density matrix $$
\gamma(\x,\x')=N\int \Psi(\x,\X) \Psi(\x',\X) d\X \ ,$$ where
$\X=(\x_2,\dots,\x_N)$ and $d\X=\prod_{j= 2}^N d^3\x_j$.

Complete (or 100\%) BEC is defined to be the property that
$\mbox{$\frac{1}{N}$}\gamma$ becomes a simple product
$f(\x)f(\x')$ as $N\to \infty$, in which case $f$ is called the {\it
condensate wave function}.  In the GP limit, i.e., $N\to\infty$
with $g=4\pi N a$ fixed, we can show that this is the case, and
the condensate wave function is, in fact, the GP minimizer
$\pgp$.

{\it THEOREM 1 (Bose-Einstein Condensation).
For each fixed $g$
$$
\lim_{N\to\infty} \frac 1 N \gamma(\x, \x') =
\pgp(\x)\pgp(\x')\ .
$$
Convergence is  in the senses that
$\Tr \left|\frac 1 N \gamma - P^{\rm GP} \right| \to 0$ and
$\int \left( \frac 1 N  \gamma(\x, \x') -
\pgp(\x)\pgp(\x')\right)^2 d^3\x d^3\x' \to 0 $.
}

We remark that Theorem 1 implies that there is 100\% condensation
for all $n$-particle reduced density matrices of $\Psi$, i.e.,
they converge to the one-dimensional projector onto the
corresponding $n$-fold product of $\pgp$. To see this, let $a^*,
a$ denote the boson creation and annihilation operators for the
state $\pgp$, and observe that
\begin{eqnarray}\nonumber
1\geq N^{-n}\langle \Psi | (a^*)^n a^n|\Psi\rangle &\approx& N^{-n}
\langle \Psi | (a^*a)^n|\Psi\rangle  \\ \nonumber
&\geq& N^{-n} \langle \Psi | a^*a|\Psi\rangle ^n \to 1 \  ,
\end{eqnarray}
where the terms coming from the commutators $[a, a^*]=1$ can be
neglected since they are of lower order as $N\to \infty$. The last
inequality follows from convexity.

Another corollary, important for the interpretation of experiments,
concerns the momentum distribution of the ground state.

{\it COROLLARY 1 (Convergence of Momentum Distribution). Let
$\widehat\rho (\k)=\int \gamma(\x, \x') \exp [i \k\cdot (\x -\x')]
 d^3\x d^3\x'$
denote the one-particle momentum  density of $\Psi$. Then, for each fixed
$g$ , $$ \lim_{N\to\infty} \frac 1N
\widehat\rho(\k)=|\widehat\phi^{\rm GP}(\k)|^2 $$ in the sense
that $\int \left|\frac 1N\widehat\rho(\k) - ~|\widehat\phi^{\rm
GP}(\k)|^2 \right|d^3\k \to 0$. Here, $\widehat\phi^{\rm GP}$
denotes the Fourier transform of $\pgp$. }

{\it Proof.} If ${\mathcal F}$ denotes the (unitary) operator `Fourier
transform' and if $\varphi$ is an arbitrary bounded function with bound
$\|\varphi\|_\infty $, then
\begin{eqnarray}\nonumber
\left|\frac 1N\int \widehat\rho\varphi-\int |\widehat\phi^{\rm
GP}|^2 \varphi\right|&=&\left|\Tr[{\mathcal F}^\dagger
(\gamma/N-P^{\rm GP}){\mathcal F}\varphi]\right|\\ \nonumber
&\leq& \|\varphi\|_\infty \Tr |\gamma/N-P^{\rm GP}|,
\end{eqnarray}
whence $\int \left|\widehat\rho/N-|\widehat\phi^{\rm
GP}|^2 \right|\leq  \Tr|\gamma/N-P^{\rm GP}|$.
\hfill QED

Before proving Theorem 1, let us state some prior results on which
we shall build. Then we shall outline the proof and  formulate
two lemmas, which will allow us to prove Theorem 1. We conclude
with the proof itself.

Denote by $\Enqm(N,a)$ the ground state energy of $H$ and by
$\Engp(g)$ the lowest energy of $\E$ with $\int |\phi|^2 =1$. The
following Theorem 2 can be deduced from \cite{LSY00}.

{\it THEOREM 2 (Asymptotics of Energy Components). Let
$\rho(\x)=\gamma(\x,\x)$ denote the density of the ground state of
$H$. For fixed $g=4\pi Na$,
\begin{mathletters}\label{partone}
\begin{equation}\label{part1a}
\lim_{N\to\infty}\frac 1N \Enqm(N,a)=\Engp(g)
\end{equation}
and
\begin{equation}\label{part1}
\lim_{N\to\infty}\frac 1N \rho(\x) = |\pgp(\x)|^2
\end{equation}
\end{mathletters}
\noindent in the same sense as in Corollary 1. Moreover, if
$\varphi_1$ denotes the solution to the scattering equation for
$v_1$ (under the boundary condition
$\lim_{|\x|\to\infty}\varphi_1(\x)=1$) and
$s=\int|\nabla\varphi_1|^2/4\pi$, then $0<s\leq 1$ and}
\begin{mathletters}\label{parttwo}
\begin{eqnarray}\nonumber
&&\lim_{N\to\infty} \int  |\nabla_{\x_1} \Psi(\x_1,\X)|^2
d^3\x_1\, d\X
\\ \label{3a}
 &&\qquad= \int|\nabla\pgp(\x)|^2d^3\x + g s \int|\pgp(\x)|^4
d^3\x,\\  &&\lim_{N\to\infty} \int  V(\x_1)|\Psi|^2 d^3\x_1\, d\X
= \int V(\x) |\pgp(\x)|^2 d^3\x, \\ \nonumber &&\lim_{N\to\infty}
\half\sum_{j=2}^N \int  v(\x_1-\x_j)|\Psi(\x_1,\X)|^2 d^3\x_1\,
d\X
\\ \label{part2} &&\qquad=(1-s) g \int|\pgp(\x)|^4 d^3\x.
\end{eqnarray}
\end{mathletters}
Only (\ref{partone}) was proved in
\cite{LSY00}, but (\ref{parttwo}) follows,
as noted in \cite{CS01a}, by multiplying $V$ and $v$ by parameters and
computing the variation of the energy with respect to  them.

(Technical note: The convergence in (\ref{part1}) was  shown in
\cite{LSY00} to be in the weak $L^1(\R^3)$ sense, but our result
here implies strong convergence, in fact. The proof in Corollary
1, together with Theorem 1 itself, implies this.)

{\it Outline of Proof:} There are two essential ingredients in our
proof of Theorem 1. The first is a proof that the part of the
kinetic energy  that is associated with the interaction $v$
(namely, the second term in (\ref{3a})) is mostly located in small
balls surrounding each particle. More precisely, these balls can
be taken to have radius  $N^{-7/17}$, which is much smaller
than the mean-particle spacing $N^{-1/3}$. This allows us to
conclude  that the function of $\x$ defined for each fixed value
of $\X$ by
\begin{equation}\label{deff}
f_\X(\x)=\frac 1{\pgp(\x)} \Psi(\x,\X)\geq 0
\end{equation}
has the property that $\nabla_\x f_\X(\x)$ is almost zero
outside the small balls centered at points of $\X$.

The complement of the small balls has a large volume but it can be
a weird set; it need not even be connected. Therefore, the
smallness of $\nabla_\x f_\X(\x)$ in this set does not guarantee
that $f_\X(\x)$ is nearly constant (in $\x$), or even that it is
continuous. We need $f_\X(\x)$ to be nearly constant in order to
conclude BEC. What saves the day is the knowledge that the total
kinetic energy of $f_\X(\x)$ (including the balls) is not huge.
The result that allows us to combine these two pieces of
information in order to deduce the almost constancy of $f_\X(\x)$
is the generalized Poincar\'e inequality in Lemma 2. {\it (End of
Outline.)}

Using the results of Theorem 2, partial integration and the GP
equation (i.e., the variational equation for $\pgp$, see
\cite{LSY00}, Eq. (2.4)) we see that
\begin{equation}\label{bound}
\lim_{N\to\infty} \int  |\pgp(\x)|^2 |\nabla_\x f_\X|^2 d^3\x\,
d\X
 = gs\int |\pgp|^4 d^3\x.
\end{equation}
The following
Lemma shows that to leading order all the energy in (\ref{bound})
is concentrated in small balls.

{\it LEMMA 1 (Localization of Energy). For fixed $\X$ let
\begin{equation}\label{defomega} \Omega_\X=\left\{\x\in \R^3
\left| \, \min_{k\geq 2}|\x-\x_k|\geq
N^{-7/17}\right\}\right. \ .
\end{equation}  Then 
$$ \lim_{N\to\infty}
\int d\X \int_{\Omega_\X} d^3\x |\pgp(\x)|^2 |\nabla_\x
f_\X(\x)|^2 = 0. $$ }

{\it Proof.} We shall show that, as $N\to \infty$,
\begin{eqnarray} \nonumber &&\int
d\X \int_{\Omega_\X^c} d^3\x\, |\pgp(\x)|^2 |\nabla_\x
f_\X(\x)|^2\\ \nonumber &&+\half\int d\X \int d^3\x\, |\pgp(\x)|^2
\sum_{k\geq 2} v(\x-\x_k) |f_\X(\x)|^2\\ \nonumber &&- 2 g \int
d\X \int d^3\x\, |\pgp(\x)|^4 |f_\X(\x)|^2\\ \label{lowbound}&&
\geq -g \int|\pgp(\x)|^4 d^3\x - o(1)\ ,
\end{eqnarray}
which implies the assertion of the Lemma by
virtue of (\ref{bound}) and the results of Theorem 2. 
Here, $\Omega_\X^c$ is the complement of $\Omega_\X$.
The proof of
(\ref{lowbound}) is actually just a detailed examination of the
lower bounds to the energy derived in \cite{LSY00} and
\cite{LY98}, and we use the methods in \cite{LSY00,LY98}, just
describing the differences from the case considered here.

Writing $f_\X(\x)=\Pi_{k\geq 2}\pgp(\x_k)F(\x,\X)$ and using that $F$
is symmetric in the particle coordinates, we see that (\ref{lowbound})
is equivalent to \begin{equation}\label{qf} \frac 1N Q(F)\geq -g
\int|\pgp|^4 - o(1), \end{equation} where $Q$ is the quadratic form
\begin{eqnarray}\nonumber &&Q(F)=\sum_{i=1}^{N} \int_{\Omega_i^c} |\nabla_i
F|^2\prod_{k=1}^{N}|\pgp(\x_k)|^2d^3\x_k\\ \nonumber
&&+\sum_{1\leq i<j\leq N} \int
v(\x_i-\x_j)|F|^2\prod_{k=1}^{N}|\pgp(\x_k)|^2d^3\x_k\\
\label{qf2} &&-2g\sum_{i=1}^{N} \int
|\pgp(\x_i)|^2|F|^2\prod_{k=1}^{N}|\pgp(\x_k)|^2d^3\x_k,
\end{eqnarray} with
$\Omega_i^c=\{(\x_1,\X)\in\R^{3N}| \, \min_{k\neq i}|\x_i-\x_k|\leq
N^{-7/17}\}$.

While  (\ref{qf}) is not true for all conceivable $F$'s satisfying
the condition $\int |F|^2\prod_{k=1}^{N}|\pgp(\x_k)|^2d^3\x_k=1$,
it {\it is} true for an $F$, such as ours, that has bounded
kinetic energy (\ref{bound}). Eqs. (4.11)--(4.12), (4.23)--(4.25),
proved in \cite{LSY00}, are similar to (\ref{qf}), (\ref{qf2}) and
almost establish (\ref{qf}), but there are two differences which
we now explain.

(i) In our case, the kinetic energy of particle $i$ is restricted
to the subset of $\R^{3N}$ in which $\min_{k\neq i}|\x_i-\x_k|\leq
N^{-7/17}$. However, from the proof of the lower bound to the ground state
energy of a homogeneous Bose gas derived in \cite{LY98} (especially Lemma
1 and Eq. (26) there), which enters the calculations in \cite{LSY00},
we see that  only this part of the kinetic energy enters the proof of
the lower bound --- except for some additional piece with a relative
magnitude  $\eps=O(N^{-2/17})$. In the notation of \cite{LY98} the radius
of the balls used in the application of Lemma 1 is chosen to be $R=
a Y^{-5/17}$, which, in the GP regime, is $R=O(N^{-7/17})$ since, for
fixed $Na$, $Y= O(a^3N) =  O(N^{-2})$. (See \cite{LY00} for a  fuller
discussion about the choice of $R$.)  The a-priori knowledge that the
total kinetic energy is bounded by (\ref{bound}) tells us that the
`additional piece', which is  $\eps$ times the total kinetic energy,
is truly $O(\eps)$ and goes to zero as $N\to \infty$.

(ii) In \cite{LSY00} all integrals were restricted to some arbitrarily
big, but finite box of size $R'$. However, the difference in the
energy is easily estimated to be smaller than $2gN\times\max_{|\x|\geq
R'}|\pgp(\x)|^2$, which, divided by $N$, is arbitrarily small, since
$\pgp(\x)$ decreases faster than exponentially at infinity (\cite{LSY00},
Lemma A.5).

Proceeding exactly as in \cite{LSY00} and taking the differences
(i) and (ii) into account, we arrive at (\ref{qf}). \hfill QED

In the following, $\K\subset\R^m$ denotes a bounded and connected
set that is sufficiently nice so that the Poincar\'e-Sobolev
inequality (see \cite{LL}, Theorem 8.12) holds on $\K$. In
particular, this is the case if $\K$ satisfies the cone property \cite{LL}
(e.g., if $\K$ is a ball or a cube).

We introduce the general notation that $f\in L^p(\K)$ if the norm
$ \|f\|_{L^p(\K)}=\left[ \int_\K |f(\x)|^p d^m\x\right]^{1/p} $ is
finite.

{\it LEMMA 2 (Generalized Poincar{\'e} Inequality). For $m\geq 2$
let $\K\subset\R^m$ be as explained above, and let $h$ be a
bounded function with $\int_\K h=1$. There exists a constant $C$
(depending only on $\K$ and $h$) such that for all sets
$\Omega\subset\K$ and all $f\in H^1(\K)$ (i.e., $f\in L^2(\K)$ and
$\nabla f\in L^2(\K)$) with $\int_\K f h\, d^m\x=0$, the inequality
\begin{eqnarray}\nonumber
&&\int_\Omega |\nabla f(\x)|^2 d^m\x
+\left(\frac{|\Omega^c|}{|\K|}\right)^{2/m}\int_\K |\nabla
f(\x)|^2 d^m\x \\ &&\geq \frac 1 C \int_{\K} |f(\x)|^2 d^m\x
 \label{poinc}
\end{eqnarray}
holds. Here $|\cdot|$ is the volume of a set, and
$\Omega^c=\K\setminus\Omega$.}

{\it Proof.} By the usual Poincar\'e-Sobolev inequality on $\K$
(see \cite{LL}, Theorem 8.12), \begin{eqnarray}\nonumber
&&\|f\|_{L^2(\K)}^2\leq \tilde C \|\nabla
f\|_{L^{2m/(m+2)}(\K)}^2\\ \nonumber &&\leq 2\tilde
C\left(\|\nabla f\|_{L^{2m/(m+2)}(\Omega)}^2+\|\nabla
f\|_{L^{2m/(m+2)}(\Omega^c)}^2\right), \end{eqnarray} if $m\geq 2$
and $\int_\K f h=0$. Applying H\"older's inequality $$ \|\nabla
f\|_{L^{2m/(m+2)}(\Omega)} \leq \|\nabla
f\|_{L^{2}(\Omega)}|\Omega|^{1/m} $$ (and the analogue with
$\Omega$ replaced by $\Omega^c$), we see that (\ref{poinc}) holds
with $C=2|\K|^{2/m}\tilde C$. \hfill QED

The important point in Lemma 2 is  that there is no restriction on
$\Omega$ concerning regularity or connectivity.

{\it Proof of Theorem 1.} For some $R>0$ let $\K=\{\x\in\R^3,
|\x|\leq R\}$, and define 
$$ \langle f_\X\rangle_\K=\frac
1{\int_\K |\pgp(\x)|^2 d^3\x} \int_\K |\pgp(\x)|^2 f_\X(\x)\, d^3\x \  .
$$ 
We shall use Lemma 2, with $m=3$,
$h(\x)=|\pgp(\x)|^2/\int_\K|\pgp|^2$, $\Omega=\Omega_\X\cap\K$ and
$f(\x)= f_\X(\x)-\langle f_\X \rangle_\K$ (see (\ref{defomega})
and (\ref{deff})). Since $\pgp$ is bounded on $\K$ above and below
by some positive constants, this Lemma also holds (with a
different constant $C'$) with $d^3\x$ replaced by $|\pgp(\x)|^2d^3\x$
in (\ref{poinc}). Therefore,
\begin{eqnarray}\nonumber
&&
\int d\X \int_\K d^3\x |\pgp(\x)|^2 \left[f_\X(\x)-\langle
f_\X\rangle_\K\right]^2
\\ \nonumber && \leq C'\int d\X\left[\int_{\Omega_\X\cap \K}
|\pgp(\x)|^2|\nabla_{\x} f_\X(\x)|^2 d^3\x\right. \\ &&\left.
\qquad\quad\qquad + \frac {N^{-8/51}}{R^2} \int_\K
|\pgp(\x)|^2|\nabla_{\x} f_\X(\x)|^2 d^3\x \right], \label{21}
\end{eqnarray}
where we used that $|\Omega_\X^c\cap\K|\leq (4\pi/3)
N^{-4/17}$. The first integral on the right side of (\ref{21})
tends to zero as $N\to\infty$ by Lemma 1, and the second is
bounded by (\ref{bound}). We conclude, since $\int_\K |\pgp(\x)|^2
f_\X(\x) d^3\x\leq \int_{\R^3} |\pgp(\x)|^2 f_\X(\x)d^3\x$, that
\begin{eqnarray}\nonumber &&\liminf_{N\to\infty} \frac 1N \langle
\pgp|\gamma|\pgp\rangle \geq \\ \nonumber &&\geq \int_\K
|\pgp(\x)|^2 d^3\x \, \lim_{N\to\infty}\int d\X \int_\K d^3\x
|\Psi(\x,\X)|^2 \  .
\end{eqnarray}
It  follows from (\ref{part1}) that the right side of this inequality
equals $\left[\int_\K |\pgp(\x)|^2 d^3\x\right]^2$.  Since the radius
of $\K$ was arbitrary, $\frac 1 N \langle\pgp|\gamma|\pgp\rangle\to 1$,
implying Theorem 1 (cf.  \cite{S79}, Theorem 2.20). \hfill QED

We remark that the method presented here also works in the case of a 2D
Bose gas. The relevant parameter to be kept fixed in the GP limit is
$g=4\pi N/|\ln (a^2 N)|$, all other considerations carry over without
essential change, using the results in \cite{LSY01,LY01}. A minor
difference concerns the parameter $s$ in Theorem 2, which can be shown
to be always equal to $1$ in 2D, i.e., the interaction energy is purely
kinetic in the GP limit (see \cite{CS01b}). We also point out that our
method necessarily fails for the 1D Bose gas, where there is presumably
no BEC \cite{PS91}. An analogue of Lemma 1 cannot hold in the 1D case
since even a hard core potential with arbitrarily small range produces an
interaction energy that is not localized on scales smaller than the total
size of the system. There is also no GP limit for the one-dimensional
Bose gas in the above sense.

We are grateful to Jakob Yngvason for helpful discussions.
E.H.L. was partially supported by the U.S. National Science Foundation
grant PHY 98 20650.  R.S. was supported by the Austrian Science Foundation
in the form of an Erwin Schr\"odinger Fellowship.

\end{document}